# Indexing of Arabic documents automatically based on lexical analysis

Abdulrahman Al_Molijy, Ismail Hmeidi, and Izzat Alsmadi

IT faculty, JUST University and Yarmouk University, Jordan

**ABSTRACT**

The continuous information explosion through the Internet and all information sources makes it necessary to perform all information processing activities automatically in quick and reliable manners. In this paper, we proposed and implemented a method to automatically create and Index for books written in Arabic language. The process depends largely on text summarization and abstraction processes to collect main topics and statements in the book. The process is developed in terms of accuracy and performance and results showed that this process can effectively replace the effort of manually indexing books and document, a process that can be very useful in all information processing and retrieval applications.

**KEYWORDS**

Text Summarization, Information Indexing, Text Classification, Text Categorization, Distance Similarity Measures, Text Stemming And Preprocessing.

## 1. INTRODUCTION

**The Arabic Language**

Arabic is a widely spoken language with more than 200 million people, distributed among many countries in the Middle East region. There are common features and characteristics for the Arabic language that may not be available in other languages. The Arabic alphabet consists of 28 letters as follow: (أ, ب, ت, ث, ج, ح, خ, د, ذ, ر, ز, س, ش, ص, ض, ط, ظ, ع, غ, ف, ق, ك, ل, م, ن, هـ, و, ي), in addition to other forms taken by some characters, such as ( والتاء المربوطة (ى), لألف المقصورة (ة), والهمزة (ء)،(ه) والهاء). Arabic letters contain also points or dots with ten characters that contain one point each (ت، ق، ي), and three letters that contain two points (ب، ج، خ، ذ، ز، ض، ظ، غ، ف، ن), and two characters that contain three points (ث,ش). Different forms of the letters appear according to their position and one word can consist of more than one part (e.g. a phrase). Letters can have different forms based on their location in the word. As letters are formed, some of which take one form, some takes two forms and some other letters take four forms. For example, the letter hamza (ء) takes only one form, and is the only character who takes one form, the letter (waw) (و, ـو) takes two forms, the letter eyen (ع), which takes one of four forms, (ع، عـ ، ـعـ ، ـع). Unlike Latin languages, Arabic text is read from right to left. Therefore, in any preprocessing to perform the following steps: Remove punctuation marks, diacritics, non-letters, and stop words.

## Documents distance similarity methods

In most natural language processing activities, measuring similarity is used for several possible reasons. For example, it is used to see the most popular words or statements and see commonalities between those words and statements. It is also used summarize text and eliminate redundancy in results. Before classification methods, document similarity methods are used to evaluate the level of similarity between a subject document and the dataset documents. Here are some examples of distance similarity metrics that are usually used in natural language processing.

## Manhattan distance

In Manhattan distance, a rank-order statistic is calculated for two profiles by measuring the difference in the positions of an N-gram in two different profiles. The general Manhattan equation is: $(pi, pj) = \sum_{h=1}^{k} |(pih - pjh)|$. The class that has the smallest Manhattan distance (i.e. the closest) is chosen as the class for the document being classified. Example, given two profiles or documents, P1, P2 consisting of two-grams where P1 = {th, he, ma, an, in, ca, ar}, and P2 = {ma, he, an, th, in, ca, ar}, then Manhattan distance (P1, P2) = (|0−3| + |1−1| + |2−0| + |3−2| + |4−4| + |5−5| + |6−6|) =6. Assume that C1 represents the profile for training class: C1 and represents the profile for training class C2, and we would like to classify a document with profile P. After computing the Manhattan distance, if Manhattan (P, C1) < Manhattan (P, C2), the document will be classified as belonging to class C1.

## Dice distance measure

The second similarity measure used is the Dice measure of similarity with the equation: (Pi, Pj)= 2|Pi ∧Pj|/|Pi| + |Pj|. Where |Pi| is the number of elements (N-grams) in profile testing and |Pi| is the number of elements (N-grams) in profile training. Using the Dice measure, the class with the largest measure is chosen as the class for the text document being classified. For Example, assume that the number of N-grams in profile Pi is |Pi| = 20, and the number of N-grams in profile Pj is |Pj|=10, and that the number of N-grams that are found in both Pi and Pj is |Pi ∧Pj| = 7, then Dice (Pi, Pj) = (2×7)/(10 + 20) = 0.467. Assume that C1 represents the profile for training class 1 and C2 represents the profile for training class 2, and we would like to classify a document with profile P. After computing the Dice measure of similarity, if Dice (P, C1) < Dice (P, C2), the document will be classified as belonging to class C2.

## N-gram classification

An N-gram is a subsequence of length N items or characters. An N-gram is like a moving window over a text, where N is the number of characters in the window. The Size of n-gram can be 1, 2 and 3 that called unigram, bigram and trigram respectively. If the size of n-gram is 4 or more, it is called an "n-gram". For example if we have the Arabic word "المسامحة", the tri-gram for this word will be: (الم , لمس , مسا , سام , امح , محة).

Following are the main tasks in N-gram classification:
First, we have modified N-gram, instead of only splitting the text into tokens consisting of three letters (tri-gram), we have also implemented algorithm with tokens of four and five letters.

To improve performance, We based classification on the first 100 words rather than the whole document.

The different options that will be evaluated then depend on preprocessing options, number of words taken (i.e. 100), number of trained documents (i.e., 1, 3 and 10, 100), and finally the N number in N-gram ( i.e. 3, 4, or 5). In every cases, the following steps must be hold: ( ., :, ; , ال , /, \, -, +, ?, <, >, @, $, %, &, *, (, ), !, ~). In every cases, the following steps must be hold:

(1) Split the text into tokens consisting only of letters.
(2) Compute all possible N-grams.
(3) Compute the frequency of occurrences for each N-gram component.
(4) Sort the N-grams according to their frequencies from most frequent to least frequent.
(5) Choose the first 100 repeated words.
(6) This gives us the N-gram profile for a document. For training class documents, the N-gram profiles were saved to a database.

## 2. RELATED WORK

Several published papers discussed the usage of different classification algorithms for Arabic text categorization. A small selection of those papers will be discussed in this section. Rehab Duwairi has several publications in this subject [5, and 14]. In [5], she discussed a survey of several published papers in text categorization and evaluated the usage of the classifiers: distance based, KNN and NB. Results showed that NB is the best classifier out of the three.

Mesleh evaluated the algorithms: SVM and several feature selection methods [6]. His results showed that CHI square showed the best performance. Baoli et al proposed an improved KNN algorithm that uses different numbers of neighbours for text categorization [7]. They conducted preliminary investigation on the Chinese language. Harrag et al proposed using neural networks for Arabic text classification [8]. They used Singular Value Decomposition (SVD) as a pre-processor method to neural networks. With several processing stages, they claimed to improve neural networks to classify Arabic text. They used a total of 453 documents in 14 classes.

Farhoodi and Yari applied machine learning algorithms for Persian text classification [9]. Examples of algorithms they used include SVM and KNN. KNN along with its cosine functions showed the best ability to detect correct classes for Persian documents. Alsaleem used also SVM and NB for automatic classification of Arabic text [10]. Author collected a large dataset of 5121 documents of 7 classes. For such large dataset, it will be interesting to see the impact of extending the classes for more than 7 classes. Based on results, SVM algorithm outperformed NB.

In a similar paper, Bawaneh et al [11] evaluated using SVM and NB for Arabic text classification. They compared their experiment with several others that were larger in datasets and training. For the 5 papers they selected, accuracy was between 71-75 to NB and 82-86 for KNN. Such results conform with most text classification papers in realizing that KNN is the best classifier. Al-Shalabi and Obeidat compared the using of KNN with N-gram classification [12]. N-gram is used as an alternative for using single words or terms in KNN. The average accuracy in case of using N-Gram is 0.7353 while with single terms indexing, it is 0.6688. In an earlier paper, Al-Shalabi et al [13] evaluated the using of KNN for Arabic text classification. However, they didn't show or compare results statically with their own or other papers.

An algorithm for Arabic text to extract the word roots is proposed by Al-Shalabi et al. The idea behind this algorithm is to assign weight and ranks to the letters that form a word based on their frequent use. The weight is a real number (between 0 to 5, Table 1). Most frequent letters get the highest weight value or 5. Other high frequent letters get: 3.5, 3, 2, and 1 respectively. The rest of the Arabic letters get 0 weight. The rank is a real number that depends on the length of the word. After determining the weights and the ranks for letters in each word, the letters with the smallest value of multiplying the ranks and the weight represent the root for the word.

Table 1. Assigning Arabic letter weights based on frequency of use

| Arabic Letters | Weight |
| --- | --- |
| ا ة | 5 |
| ي ىء | 3.5 |
| ت ى و | 3 |
| أ إ م ن | 2 |
| ل س ه | 1 |
| Rest of the Arabic Alphabet | 0 |

Similar to earlier papers, Thabtah et al evaluated using Chi square, VSM and KNN and NB for Arabic text classification [15 and 16].

## 3. GOALS AND APPROACHES

The research and practical contribution in automatic Arabic indexing is minimal. The number of work in English document indexing is much more in contrast with Arabic document indexing. The goal is to build an efficient algorithm to perform automatic Arabic indexing with a high level of accuracy. As discussed earlier, the nature of Arabic text is different from the English text, so the preprocessing of the Arabic text will be different and a little bit more challenging. Before applying the indexing process, the documents should be preprocessed. The preprocess phase includes removing irrelevant text that can be found in the text. This may include: punctuation marks, diacritics, non-letters, stop words and any words that are not part of the main document language (e.g. English words when the document is largely written in Arabic). The rest of the document is retrieved after removing irrelevant text. The number of these words can be large. In addition ranking techniques are implemented to select the most important terms. After that, to increase the accuracy of the indexing process, we should extract the roots of the words in order to reduce the number of the words. There are many methods to extract the root for the word in the Arabic text (e.g. ref [3]). We will apply a modified version of this method to extract roots of words in Arabic.

**Dataset Description**

A dataset of Arabic documents is collected from several online Arabic websites such as: (www.star28.com and www.kutub.info). Each document file size is between 300 to 800 KB and contains between 1500 to 4000 words. 50 documents from 10 different categories are selected. The categories of the documents include: Banks, commerce, education, crimes, religion, sport, tourism, food, sciences, women, are Computer Network and Technology.

**Inverted index**

Inverted index is an index data structure that saves a mapping from each term in the document to its locations in a document or a set of documents. The purpose of an inverted index is to allow fast full text searches, at a cost of increased processing when a document is added to the database. The inverted file may be the database file itself, rather than its index. It is the most popular data structure used in document or information retrieval systems such as in search engines.

There are two main variants of inverted indexes: An inverted file index that contains a list of references to documents for each term. A word level inverted index (or full inverted index or inverted list) additionally contains the position of each word within the document. The latter form offers more functionality (e.g. phrase searches). However, it needs more time and space to be created.

**Indexing methods applied**

The method that we choose to solve the first part of the system (Automatic Arabic document indexing problem) is as follow.

The indexing process starts from preprocessing the document by removing irrelevant text such as: punctuation marks, diacritics, non-letters, stop words and any English word or letter.

Then the system computes the frequency of every term in the document and reorder them decreasing.

Using ranking techniques will remove all term with highest and lowest frequency.
After that, the system matches between the term and the page number where the term occurs in the document.

Finally, we add the index to the end of the document.
For the second part of the system (i.e. the inverted index), the process is similar to the first or the indexing stage with some differences. The developed program first check if the document exists in the already database of documents. If not, the system starts to preprocess the document and remove all irrelevant text. Then the system computes the frequency of every term in the document and reorder them decreasing.

**Comparison between manual and automatic indexing**

In principle, manual indexing is considered as a reference and more accurate than automatic indexing. This is largely since humans are more capable of dealing with the semantic complexity of natural language specially in cases where the same word may have different meanings based on the sentence and paragraph context. Table 1 summarizes the major differences between manual and automatic document indexing.

Table 1: differences between human and automatic indexing.

| Variables | Human Indexing | Automatic Indexing |
| --- | --- | --- |
| Cost | expensive per unit indexed because it is labor-intensive | inexpensive per unit indexed |
| Time | involves more time per unit indexed | can index large amounts of material in short amount of time |
| Extent of indexable matter | may be limited to abstract or summarization of text | routinely based on complete text |
| Exhaustively | tends to be more selective | considers most of the words in indexable material |
| Specificity | uses more generic terminology, smaller vocabulary | uses very specific terminology, larger vocabulary |
| Browsable displayed indexes | use multi-term context-providing headings | limited use of term combinations |
| Searching syntax, display syntax | use wide-range of syntactic patterns and can adapt quickly to include newer terminology, as well as older subject headings | Becoming more sophisticated, and is selecting, combining, manipulating, and weighing terms. Usually limited to key-words in, out of, or along-side context. |
| Vocabulary management | can cross-reference, link synonyms or like terms, point to related terms easily | being experimented with |
| Surrogation | not often used by human indexing | being used frequently –often as visual displays, such as icons or graphs |

**Evaluation and Analysis**

Based on the different criteria described earlier, experiments are conducted to compare results from automatic indexing relative to manual indexing. Two performance metrics are considered: Precision and recall. They are calculated according to the following formulas.

$$\text{Precision} = \frac{tp}{tp + fp} \quad \text{Recall} = \frac{tp}{tp + fn}$$

Figure 1. Prediction performance metrics.

The overall average precision for the 5 documents was 0.998 while recall was 1.00. This simply means that there was a few values of FP (False Positive), and no value in FN (False Negative). Accuracy of the results was very high and this is possibly since the size of the dataset is relatively small.

## 4. SUMMARY AND CONCLUSION

The large amount of Arabic documents available through the Internet requires the need for an efficient way to index them for search and retrieval. This paper presents an automatic Arabic document indexing method. The indexing process starts by implementing preprocessing steps to remove irrelevant text (e.g. punctuation marks, diacritics, non-letters, etc.). Then the system computes the frequency of every term in the document and reorders them in a descending order. A ranking algorithm is then used to remove all terms with highest and lowest frequency. After that, the system matches between the term and the page number where the term occurs in the document and automatically adds the index to the end of the document.

The developed program is developed based on accuracy and performance and results showed that it can effectively replace the human time consuming effort for indexing a large number of documents or books. Such process can be used and useful for several information retrieval and processing applications.

International Journal on Natural Language Computing (IJNLC) Vol. 1, No.1, April 2012